\documentclass[aps,reprint,twocolumn,citeautoscript,showpacs,amsmath,amssymb,prl]{revtex4-1}
\usepackage{graphicx}
\usepackage{dcolumn}
\usepackage{bm}
\usepackage{times}
\usepackage{color}
\begin{document}
\title{The non-magnetic collapsed tetragonal phase of CaFe$_2$As$_2$ and superconductivity in the iron pnictides}
\author{J. H. Soh,$^{1}$ G. S. Tucker,$^{1}$ D. K. Pratt,$^{2}$ D. L. Abernathy,$^{3}$ M. B. Stone,$^{3}$ S. Ran,$^{1}$ S.~L. Bud$^{\prime}$ko,$^{1}$ P. C. Canfield,$^{1}$ A. Kreyssig,$^{1}$ R. J. McQueeney$^{1}$
 and A. I. Goldman$^{1}$}
\affiliation{$^1$Ames Laboratory, U.S. DOE and Department of Physics and Astronomy, Iowa State University, Ames, Iowa 50011, USA}
\affiliation{$^2$NIST Center for Neutron Research, National Institute of Standards and Technology, Gaithersburg, Maryland 20899, USA}
\affiliation{$^3$Quantum Condensed Matter Division, Oak Ridge National Laboratory, Oak Ridge, Tennessee 37831, USA}

\date{\today}

\begin{abstract}
The relationship between antiferromagnetic spin fluctuations and superconductivity has become a central topic of research in studies of superconductivity in the iron pnictides. We present unambiguous evidence of the absence of magnetic fluctuations in the non-superconducting collapsed tetragonal phase of CaFe$_2$As$_2$ via inelastic neutron scattering time-of-flight data, which is consistent with the view that spin fluctuations are a necessary ingredient for unconventional superconductivity in the iron pnictides. We demonstrate that the collapsed tetragonal phase of CaFe$_2$As$_2$ is non-magnetic, and discuss this result in light of recent reports of high-temperature superconductivity in the collapsed tetragonal phase of closely related compounds.
\end{abstract}

\maketitle

The $A$Fe$_2$As$_2$ ($A$ = Ba, Sr, Ca), or "122", family of compounds has been one of the most widely studied classes of iron pnictide superconductors \cite{Johnston_2010,PandG_2010,CandB_2010,Stewart_2011} in recent years, and a great deal of attention has been focused on CaFe$_2$As$_2$ \cite{Ni_2008,PC_2009} in particular.  At ambient pressure, the substitution of Co or Rh for Fe \cite{Kumar_2009,Matusiak_2010,Harnagea_2011,Ran_2012,Danura_2011} results in the suppression of antiferromagnetic (AFM) order and, over some range in substitution, superconductivity (SC) emerges with transition temperatures ($T_{\rm{c}}$) of up to $\approx$~20~K. Under modest applied pressure Ca(122) manifests fascinating new behavior including a transition to an isostructural volume collapsed tetragonal (cT) phase that is generally believed to be non-magnetic and non-superconducting. The cT phase in Ca(122) is distinguished by a striking 9.5\% reduction in the tetragonal \textbf{c} lattice parameter, with respect to the high-temperature ambient-pressure tetragonal (T) phase, along with the absence of the stripe-like magnetic order found for the low-temperature ambient-pressure orthorhombic phase \cite{Goldman_2008}.

The first liquid media clamp-cell pressure measurements of Sn-flux solution-grown Ca(122) found traces of SC for applied pressures between roughly 0.25 and 0.9 GPa \cite{Milton_2008,Park_2008}. These studies were rapidly followed by transport measurements and neutron diffraction experiments under hydrostatic pressure conditions using He gas pressure cells which showed: (i) no evidence of SC for P $< 0.6$ GPa \cite{Yu_2009} and; (ii) the existence of the cT structure for $P >$ 0.35 GPa at low temperatures \cite{Kreyssig_2008,Goldman_2009}. That work demonstrated that the traces of SC originally found in the frozen liquid clamp-cell measurements probably resulted from significant non-hydrostatic pressure components generated during the transition to the cT phase, although the origin of the SC phase was not identified in these studies.  Later experiments, utilizing uniaxial pressure, concluded that the T phase could be stabilized to low temperatures by the presence of non-hydrostatic pressure components and was likely the source of superconductivity in the original liquid clamp-cell measurements \cite{Prokes_2010}.

Recently, superconductivity with $T_{\rm{c}}$ in excess of 45~K has been reported for the substitution of Sr \cite{Jeffries_2012} or selected rare earths ($R$) \cite{Lv_2011, Saha_2012,Ma_2013} for Ca, or co-doping by La and P \cite{Kudo_2013}, and it has been proposed that these high $T_{\rm{c}}$ values are realized in the cT phase as well \cite{Saha_2012,Jeffries_2012}. Since it is generally accepted that there is a close connection between SC in the iron pnictides and the presence of correlated AFM fluctuations in these compounds \cite{Johnston_2010,PandG_2010,CandB_2010,Stewart_2011,LandC_2010,Dai_2012}, the possibility of high values of $T_{\rm{c}}$ in the cT phase raises important questions regarding the nature of the cT phase, and the relationship between magnetic fluctuations and unconventional superconductivity in the iron pnictides.  It is, therefore, important to clearly establish whether the cT phase of Ca(122) is, in fact, non-magnetic.

There is already evidence that the cT phase of Ca(122) is non-magnetic, consistent with the absence of unconventional superconductivity.  First, as noted above, the low-temperature stripe-like AFM order is absent in the cT phase. However, alternative magnetic ground states for the cT phase have been proposed \cite{Yildirim_2009}, and the origin of the suppression of magnetic order, whether it arises from a reduction in the iron moment, changes in the magnetic exchange, or a more subtle change in electronic structure has come under renewed scrutiny \cite{Jeffries_2012}.  Furthermore, the absence of AFM order does not directly speak to the presence or absence of magnetic \emph{fluctuations} in the cT phase. It is well known that strong AFM fluctuations remain after long-range magnetic order is lost in the iron pnictides at optimal doping \cite{Johnston_2010,PandG_2010,CandB_2010,Stewart_2011,LandC_2010,Dai_2012}.

Total energy calculations described in Reference~\onlinecite{Kreyssig_2008} predict that the cT phase is non-magnetic and this has been supported by other theoretical studies \cite{Colonna_2011,Tomic_2012,Widom_2013}.   Our previous inelastic neutron scattering studies of the T \cite{Diallo_2010} and cT phases \cite{Pratt_2009} showed that, at least over a narrow range in momentum transfer (\textbf{Q}) close to the AFM wavevector, $\textbf{Q}_{\rm{stripe}}$, and energy transfers ($E$) less than 7 meV, the AFM fluctuations are suppressed, or absent, in the cT phase. Again, this result finds support in other experimental measurements \cite{Danura_2011,Ma_2013}. But the narrow scope of the neutron measurements could not exclude the presence of correlated magnetic fluctuations at other positions in reciprocal space \cite{Yildirim_2009}, or simply a change in the energy scale of the fluctuations as has been found, for example, in the well known volume collapse of Ce \cite{Loong_1987}, or very recently in nonsuperconducting Ba(Fe$_{0.85}$Ni$_{0.15}$)$_2$As$_2$ \cite{Wang_2013}. A much wider view in both \textbf{Q} and $E$ must be obtained to clearly establish the presence or absence of magnetic fluctuations in the cT phase of Ca(122).

Here we present unambiguous evidence that the magnetic fluctuations in the non-superconducting cT phase of Ca(122) are absent via inelastic neutron scattering measurements using the ARCS time-of-flight (TOF) instrument \cite{Abernathy_2012} at the Spallation Neutron Source at Oak Ridge National Laboratory.  This result provides clear evidence that the cT phase of Ca(122) is a non-magnetic metal, with no static or dynamic magnetic moment, and supports the view that spin fluctuations are a necessary ingredient for unconventional SC in the iron pnictides.  The complete suppression of magnetism in the cT phase also provides a non-magnetic analog for a detailed study of the AFM fluctuation spectrum of the paramagnetic T phase out to energy transfers above 100 meV, and we use this to demonstrate that the dynamical susceptibility, $\chi^{\prime\prime}(\textbf{\rm{Q}},\omega)$, is well described by the model for short-range, over-damped anisotropic spin-correlations introduced in Reference~\onlinecite{Diallo_2010}.

The sample used in this study was a co-aligned set of 12 single crystals produced by solution growth using an FeAs flux \cite{Ran_2011}.  The co-alignment provided a total sample mass of $\sim$1.5 grams and a sample mosaic of 1.5$^{\circ}$ full-width-at-half-maximum.  As described in Reference~\onlinecite{Ran_2011}, FeAs-flux samples quenched from the melt at 960$^{\circ}$C, or annealed at temperatures above 700$^{\circ}$~C, transform directly from the T phase into the cT structure at low temperature at ambient pressure; the strain field associated with a uniform distribution of fine-sized FeAs precipitates appears to play a key role in the ambient pressure transformation and can be used to systematically tune the behavior of the Ca(122) samples \cite{Gati_2012}. For the present measurements, the samples were as-grown, quenched from the melt at 960$^{\circ}$~C. Other than a shift in temperature, the transformation from the T phase to the cT phase at ambient pressure is consistent with the T-cT transformation observed for the Sn-flux solution-grown samples under applied pressure \cite{Ran_2011}, eliminating the need for a pressure cell and, therefore, the dominant contribution it makes to the measured background in scattering measurements.

The inelastic neutron scattering experiment was performed using incident beam energies of 75 meV and 250 meV.  The sample was attached to the cold-finger of a closed-cycle cryostat and oriented with the tetragonal \textbf{c}-axis parallel to the incident beam. In what follows, the neutron scattering data will be described in the tetragonal $I4/mmm$ coordinate system with $\textbf{Q} = \frac{2\pi}{a}(H + K)\hat{\imath} + \frac{2\pi}{a}(H - K)\hat{\jmath} + \frac{2\pi}{c}L\hat{k} = (H + K,H - K,L)$. In this notation, the stripe-like AFM wavevector is $\textbf{Q}_{\rm{stripe}} = (\frac{1}{2},\frac{1}{2},1)$ [$H = \frac{1}{2}, K = 0$].  $H$ and $K$ are defined to conveniently describe diagonal cuts in the $I4/mmm$ basal plane as varying $H$ ($K$) corresponds to a longitudinal [$H,H$] scan (transverse [$K,-K$] scan) through $\textbf{Q}_{\rm{stripe}}$. It can also be shown that $H$ and $K$ are the reciprocal lattice vectors of the Fe square lattice as discussed in Ref.~\onlinecite{Tucker_2012}.

\begin{figure}
\centering\includegraphics[width=1
\linewidth]{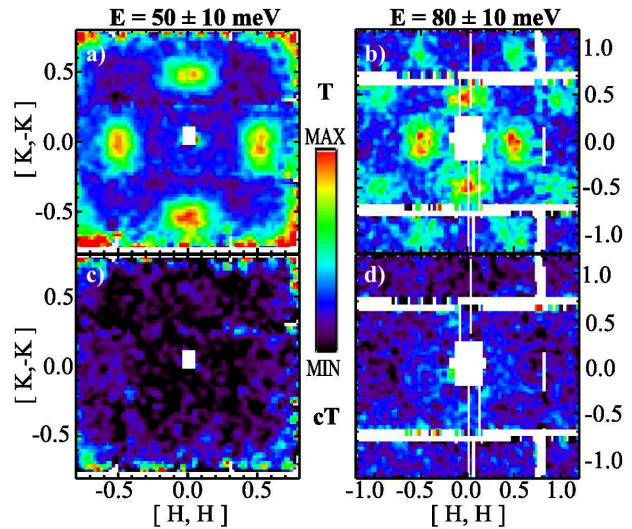}\\
\caption{(color online) Constant energy ($\Delta$$E$ = $\pm$10 meV) slices in the [$H,H$] - [$K,-K$] plane with incident beam energies of 75 meV (left panel) and 250 meV (right panel). The color scale shows the intensity of the scattered neutrons in the stated energy intervals. The white regions correspond to gaps in the detector coverage. The paramagnetic T phase at 150~K is shown in panels (a) and (b). There is no evidence of magnetic excitations in the cT phase at 10 K in panels (c) and (d). All panels employ the same intensity scale.}\label{figure1}
\end{figure}

\begin{figure}
\centering\includegraphics[width=1
\linewidth]{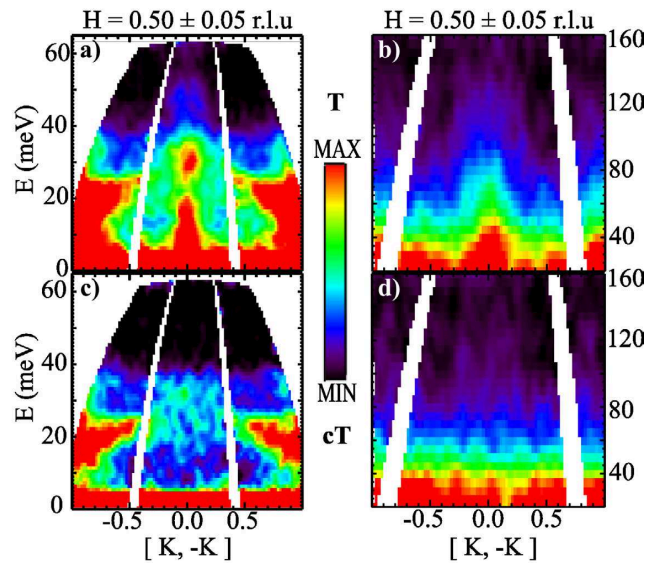}\\
\caption{(color online) Magnetic excitation spectra along the [$K,-K$] direction, averaging over an interval of $\pm 0.05$ r.l.u. in the [$H,H$] direction for incident energies of 75 meV (left) and 250 meV (right). In panels (a) and (b) the magnetic excitations are clearly evident as plumes of intensity around K = 0 in the paramagnetic T phase at 150~K. At low temperature, in the cT phase at 10 K [panels (c) and (d)], magnetic excitations are absent.  The dispersing features entering from the left and right sides of panels (a) and (c) are phonons from aluminum. All panels employ the same intensity scale.}\label{figure2}
\end{figure}

We performed a detailed survey of the spin fluctuations at temperatures above ($T$ = 150~K) and below ($T$ = 10~K) the T-cT transition (at $\approx$ 90~K) and used the MSLICE software \cite{Coldea_2004} to visualize the data and to take one and two-dimensional cuts through main crystallographic symmetry directions for subsequent data analysis.  Figures~\ref{figure1} and \ref{figure2} display the key result of our measurements. Figures~\ref{figure1}~(a) and (b) show the neutron intensity for constant energy slices (integrated over $\Delta$$E$ = $\pm 10$ meV) for $E_{\rm{i}}$ = 75 meV and $E$ = 50 meV [Fig.~~\ref{figure1}~(a)], and $E_{\rm{i}}$ = 250 meV and $E$ = 80 meV [Fig.~~\ref{figure1}~(b)] taken at 150~K, above the T-cT transition.  The AFM spin fluctuations centered at $\textbf{Q}_{\rm{stripe}}$ and equivalent positions in other Brillouin zones (for $E_{\rm{i}}$ = 250 meV) are clearly observed.  Figures~\ref{figure1}~(c) and (d) show the neutron intensity for these same energy slices taken at $T$ = 10~K in the cT phase, demonstrating the absence of magnetic scattering in the vicinity of $\textbf{Q}_{\rm{stripe}}$, and we find no evidence of magnetic intensity at other positions in reciprocal space.

\begin{figure}
\centering\includegraphics[width=1
\linewidth]{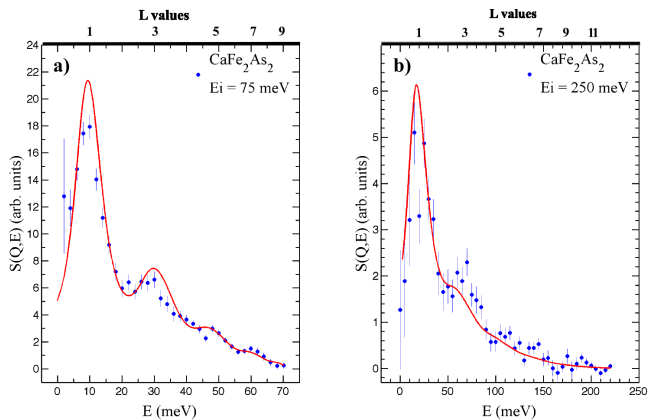}\\
\caption{(color online) The dynamical structure factor, $S(\textbf{\rm{Q}},\rm{E})$, as a function of energy measured at $\textbf{Q}_{\rm{stripe}}$ = (1/2 1/2 1) for incident neutron energies of a) 75 meV and b) 250 meV. The scales at the top show the relationship between $L$ and E. The solid lines represent a model fit to the data as described in the text.}\label{figure3}
\end{figure}

\begin{figure}
\centering\includegraphics[width=0.95
\linewidth]{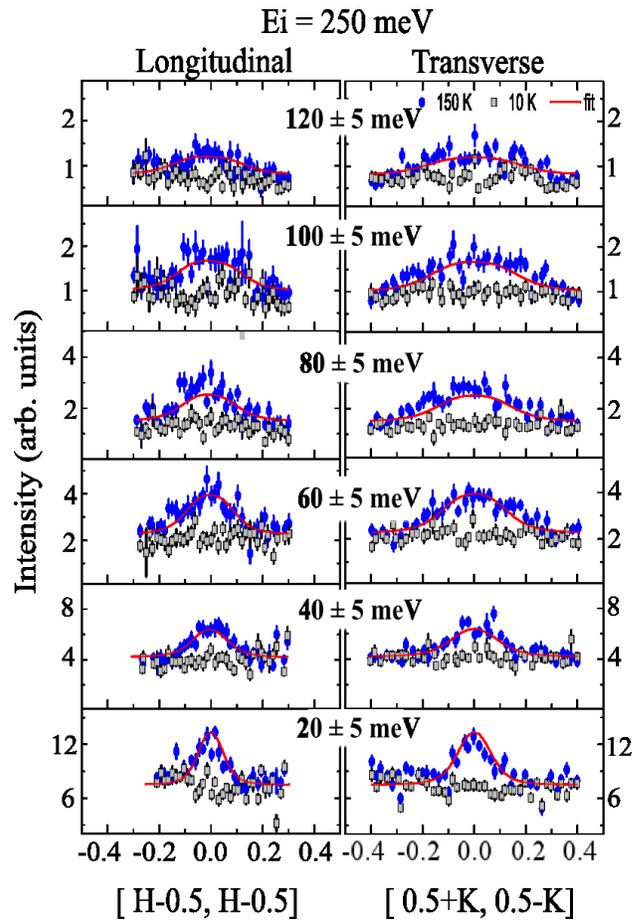}\\
\caption{(color online) Constant-energy \textbf{Q}-cuts of the magnetic scattering along the longitudinal [$H,H$] direction (left panel) and transverse [$K,-K$] direction (right panel) with an incident neutron energy of 250 meV for the T phase at 150~K (blue circles) and the cT phase at 10~K (shaded squares). The in-plane anisotropy of the scattering is evident from the difference in the widths of the scattering between the longitudinal and transverse directions at a given energy transfer.  The solid lines represent fits to the data as described in the text.}\label{figure4}
\end{figure}

Figure~\ref{figure2} shows the energy dependence of the magnetic intensity along the [$K,-K$] direction after averaging over the longitudinal [$H,H$] direction from 0.45 $< H <$ 0.55 in reciprocal lattice units (r.l.u).  Figures~\ref{figure2}~(a) and (b) show the neutron intensity for $E_{\rm{i}}$ = 75 meV and 250 meV, respectively, taken at $T$ = 150~K.  In the T phase, the plume of scattering at $\textbf{Q}_{\rm{stripe}}$ extends above 100 meV [Fig.~\ref{figure2}~(b)]. The data taken in the cT phase, at $T$ = 10~K, again show no evidence of magnetic scattering in this region (see also Fig.~\ref{figure4}).  Taken together, Figs.~\ref{figure1} and \ref{figure2} clearly demonstrate that AFM fluctuations are absent in the cT structure consistent with the absence of any Fe moment whatsoever.

The full suppression of magnetism and the absence of SC in the cT phase supports current theories of unconventional pairing in the iron pnictides via spin fluctuations, and raises important questions regarding the origin of SC in the cT phase of (Ca$_{1-x}$Sr$_x$)Fe$_2$As$_2$ with $T_{\rm{c}} \simeq 22$~K \cite{Jeffries_2012}, and (Ca$_{1-x}R_x$)Fe$_2$As$_2$ ($R$ = Pr, Nd) with $T_{\rm{c}} > 45$~K \cite{Saha_2012}.  Both References \onlinecite{Jeffries_2012} and \onlinecite{Saha_2012} acknowledge the possibility of the SC originating in a second phase, perhaps within some retained T phase as found for CaFe$_2$As$_2$ under uniaxial pressure \cite{Prokes_2010}. On the other hand, the values for $T_{\rm{c}}$ in these systems is significantly higher than that found for CaFe$_2$As$_2$ ($\approx$ 10~K), offering the possibility that SC arises from an alternative pairing scenario.  Clearly, it would be instructive to study examples of the Sr and $R$-substituted compounds using the TOF methods described here in order to establish whether remnants of magnetic fluctuations persist into the cT phase.

The absence of magnetic scattering in the cT phase provides us with a non-magnetic analog to serve as a background reference for a detailed investigation of spin fluctuations in the paramagnetic T phase.  Figure~\ref{figure3} displays the energy spectrum for the spin fluctuations in the T phase for both incident neutron energies.  These plots were obtained from a subtraction of the data obtained at 150~K and 10~K, then folding the resultant difference spectrum across the diagonals of Fig.~\ref{figure1}. This folding effectively increases the statistics by taking advantage of the fourfold symmetry of the [$H,H$]$-$[$K,\rm{-}K$] plane. We note that no adjustment of the data to account for the temperature factor was done in the subtraction because it was not possible to assign relative weights to the temperature dependent (e.g. sample, sample holder) and independent (scattering from the cryostat, general background) contributions to the cT data with any certainty. Nevertheless, the absence of a correction for the temperature factor in the subtraction affects only energies below approximately 15 meV and is of no consequence for the analysis described below. The range of integration over \textbf{Q} in Fig.~\ref{figure3} was $\Delta$$H$ = 0.45 to 0.55 r.l.u. and $\Delta$$K$ = -0.06 to +0.06 r.l.u., for consistency with Ref.~\onlinecite{Diallo_2010}.  The intensity modulation with energy arises from variations in the structure factor along $L$ which are observed as energy-dependent intensity oscillations that are peaked at the AFM zone centers (e.g. $L$ = 1, 3, 5).\cite{Diallo_2010}.

Complementing these data, in Fig.~\ref{figure4} we show constant-energy cuts through $\textbf{Q}_{\rm{stripe}}$ along the longitudinal [$H,H$] and transverse [$K,-K$] directions for energy transfers from 20 to 120 meV.  Data taken in the paramagnetic T phase at 150~K (blue circles) are contrasted with the corresponding cuts in the cT phase at 10~K (shaded squares), once again demonstrating the absence of any magnetic signal in the cT phase.  Furthermore, the background scattering away from $\textbf{Q}_{\rm{stripe}}$ in the T phase is indistinguishable from the scattering in the cT phase indicating that there is no additional incoherent paramagnetic contribution.

Following Ref.~\onlinecite{Diallo_2010}, the \textbf{Q} and constant-energy cuts in Figs.~\ref{figure3} and \ref{figure4} can be described by a scattering model that includes short-range and anisotropic spin correlations with overdamped dynamics.  The dynamic susceptibility can be written as:
\begin{widetext}
\begin{equation}\label{eqn1}
   \chi^{\prime\prime}(\textbf{\rm{Q}},\omega)=\frac{\hbar\omega\gamma\chi_0}{(\hbar\omega)^2 + \gamma^2\{(q^2+\eta q_xq_y)a^2 + (\frac{\xi_T}{a})^{-2} + \eta_c[1 + \cos(\pi L)]\}^2}
\end{equation}
\end{widetext}
where $q^2 = q_{x}^{2} + q_{y}^{2}$, $\chi_0$ is the staggered susceptibility, $\gamma$ denotes the damping coefficient originating from the spin decay into particle-hole excitations, and $\xi_T$ and $a$ are the magnetic correlation length at temperature $T$, and the in-plane lattice parameter, respectively.  Two dimensionless parameters describe the anisotropy of the in-plane correlation lengths ($\eta$) and the strength of the interlayer spin correlations ($\eta_c = J_c\chi_0$).

The dynamical structure factor, $S(\textbf{\rm{Q}},\omega)$ is related to $\chi^{\prime\prime}(\textbf{\rm{Q}},\omega)$ by the fluctuation-dissipation theorem, so that:
\begin{equation}\label{eqn2}
    S(\textbf{\rm{Q}},\omega) = CF(\textbf{\rm{Q}})^2\frac{\chi^{\prime\prime}(\textbf{\rm{Q}},\omega)}{1-e^{-\hbar\omega/kT}}
\end{equation}
where $F(\textbf{\rm{Q}})$ is the Fe$^{2+}$ magnetic form factor, $C$ is a scaling constant and E = $\hbar\omega$.  Fits to the energy spectrum (solid line in Fig.~\ref{figure3}) and constant-energy \textbf{Q}-cuts (solid lines in Fig.~\ref{figure4}) were performed simultaneously using Eqns.~\ref{eqn1} and \ref{eqn2} and a single scale factor for each incident energy. We obtained values for $\gamma$ = 37$\pm$2 meV, $\xi_T$ = 6.4$\pm$0.2 {\AA}, $\eta$ = 1.0$\pm$0.2 and $\eta_c$ = 0.16$\pm$0.02 that compare well with those determined for the paramagnetic T phase at 180~K in Ref.~\onlinecite{Diallo_2010}: $\gamma$ = 43$\pm$5 meV, $\xi_T$ = 7.9$\pm$0.1 {\AA}, $\eta$ = 0.55$\pm$0.36 and $\eta_c$ = 0.20$\pm$0.02. We point out here that the present data set extends to much higher energies than previously measured for Ca(122) and, therefore, provides further validation of the nearly AFM spin fluctuation model proposed by Diallo \emph{et al.} \cite{Diallo_2010} and, in addition, shows that the spin dynamics of the FeAs flux-grown samples and the original Sn flux-grown samples are the same.

In summary, our inelastic neutron scattering data, over an extended range in reciprocal space and energy, demonstrate that the cT phase of Ca(122) is non-magnetic.  Based on an accurate background subtraction using the non-magnetic cT phase, we find no evidence for spin fluctuations at other wave vectors, or any incoherent contribution, and conclude that the magnetic fluctuations are exclusive to $\textbf{Q}_{\rm{stripe}}$ for energies below 120 meV.  In light of recent reports of high-temperature SC in the cT phase of (Ca$_{1-x}$Sr$_x$)Fe$_2$As$_2$ \cite{Jeffries_2012} and (Ca$_{1-x}R_x$)Fe$_2$As$_2$ ($R$ = Pr, Nd) \cite{Saha_2012} the absence of spin fluctuations in the cT phase of Ca(122) clearly calls for further consideration of multiple phases as the source of SC in these systems as well as similar neutron TOF measurements on these compounds.

\begin{acknowledgments}
Work at the Ames Laboratory was supported by the Department of Energy, Basic Energy Sciences under Contract No. DE-AC02-07CH11358. A portion of this research at ORNL's Spallation Neutron Source was sponsored by the Scientific User Facilities Division, Office of Basic Energy Sciences, U.S. Department of Energy.
\end{acknowledgments}


\begin{thebibliography}{99}

\bibitem{Johnston_2010} D. C. Johnston, Adv. Phys. \textbf{59}, 803 (2010).
\bibitem{PandG_2010} J. Paglione and R. L. Greene, Nat. Phys. \textbf{6}, 645 (2010).
\bibitem{CandB_2010} P. C. Canfield and S. L. Bud'ko, Annu. Rev. Condens. Matter Phys. \textbf{1}, 27 (2010).
\bibitem{Stewart_2011} G. R. Stewart, Rev. Mod. Phys. \textbf{83}, 1589 (2011).
\bibitem{Ni_2008} N. Ni, S. Nandi, A. Kreyssig, A. I. Goldman, E. D. Mun, S. L. Bud'ko, and P. C. Canfield, Phys. Rev. B \textbf{78}, 014523 (2008).
\bibitem{PC_2009} P. C. Canfield, S. L. Bud'ko, N. Ni, A. Kreyssig, A. I. Goldman, R. J. McQueeney, M. S. Torikachvilli, D. N. Argyriou, G. Luke and W. Yu, Physica C \textbf{469}, 404 (2009).
\bibitem{Kumar_2009} Neeraj Kumar, R. Nagalakshmi, R. Kulkarni, P. L. Paulose, A. K. Nigam, S. K. Dhar, and A. Thamizhavel, Phys. Rev. B \textbf{79}, 012504 (2009).
\bibitem{Matusiak_2010} M. Matusiak, Z. Bukowski, and J. Karpinski, Phys. Rev. B \textbf{81}, 020510(R) (2010).
\bibitem{Harnagea_2011} L. Harnagea, S. Singh, G. Friemel, N. Leps, D. Bombor,M. Abdel-Hafiez, A. U. B.Wolter, C. Hess, R. Klingeler, G. Behr, S.Wurmehl, and B. B\"{u}chner, Phys. Rev. B \textbf{83}, 094523 (2011).
\bibitem{Ran_2012} S. Ran, S. L. Bud'ko, W. E. Straszheim, J. Soh, M. G. Kim, A. Kreyssig, A. I. Goldman, and P. C. Canfield, Phys. Rev. B \textbf{85}, 224528 (2012).
\bibitem{Danura_2011} M. Danura, K. Kudo, Y. Oshiro, S. Araki, T. C. Kobayashi and M. Nohara, J. Phys. Soc. Jpn. \textbf{80}, 103701 (2011).
\bibitem{Goldman_2008} A. I. Goldman, D. N. Argyriou, B. Ouladdiaf, T. Chatterji, A. Kreyssig, S. Nandi, N. Ni, S. L. Bud'ko, P. C. Canfield, and R. J. McQueeney, Phys. Rev. B \textbf{78}, 100506(R) (2008).
\bibitem{Milton_2008} M. S. Torikachvili, S. L. Bud'ko, N. Ni, and P. C. Canfield, Phys. Rev. Lett. \textbf{101}, 057006 (2008).
\bibitem{Park_2008} T. Park, E. Park, H. Lee, T. Klimczuk, E. D. Bauer, F. Ronning, and J. D. Thompson, J. Phys.: Condens. Matter \textbf{20}, 322204 (2008).
\bibitem{Yu_2009} W. Yu, A. A. Aczel, T. J. Williams, S. L. Bud'ko, N. Ni, P. C. Canfield, and G. M. Luke, Phys. Rev. B \textbf{79}, 020511(R) (2009).
\bibitem{Kreyssig_2008} A. Kreyssig, M. A. Green, Y. B. Lee, G. D. Samolyuk, P. Zajdel, J. W. Lynn, S. L. Bud'ko, M. S. Torikachvili, N. Ni, S. Nandi, J. B. Le{\~ a}o, S. J. Poulton, D. N. Argyriou, B. N. Harmon, R. J. McQueeney, P. C. Canfield, and A. I. Goldman, Phys. Rev. B \textbf{78}, 184517 (2008).
\bibitem{Goldman_2009} A. I. Goldman, A. Kreyssig, K. Proke{\v s}, D. K. Pratt, D. N. Argyriou, J. W. Lynn, S. Nandi, S. A. J. Kimber, Y. Chen, Y. B. Lee, G. D. Samolyuk, J. B. Le\~{a}o, S. J. Poulton, S. L. Bud$'$ko, N. Ni, P. C. Canfield, B. N. Harmon, and R. J. McQueeney, Phys. Rev. B \textbf{79}, 024513 (2009).
\bibitem{Prokes_2010} K. Proke{\v s}, A. Kreyssig, B. Ouladdiaf, D. K. Pratt, N. Ni, S. L. Bud'ko, P. C. Canfield, R. J. McQueeney, D. N. Argyriou, and A. I. Goldman, Phys. Rev. B \textbf{81}, 180506(R) (2010).
\bibitem{Jeffries_2012} J. R. Jeffries, N. P. Butch, K. Kirshenbaum, S. R. Saha, G. Samudrala, S. T. Weir, Y. K. Vohra, and J. Paglione, Phys. Rev. B \textbf{85}, 184501 (2012).
\bibitem{Lv_2011} B. Lv, L. Deng, M. Gooch, F. Wei, Y. Sun, J. K. Meen, Y.-Y. Xue, B. Lorenz and C.-W. Chu, Proc. Nat. Acad. Sci. \textbf{108}, 15705 (2011).
\bibitem{Saha_2012} S. R. Saha, N. P. Butch, T. Drye, J. Magill, S. Ziemak, K. Kirshenbaum, P. Y. Zavalij, J. W. Lynn, and J. Paglione, Phys. Rev. B \textbf{85}, 024525 (2012).
\bibitem{Ma_2013} L. Ma, G.-F. Ji, J. Dai, S. R. Saha, T. Drye, J. Paglione, W.-Q. Yu, Chin. Phys. B \textbf{22}, 057401 (2013).
\bibitem{Kudo_2013} K. Kudo, K. Iba, M. Takasuga, Y. Kitahama, J. Matsumura, M. Danura, Y. Nogami and M. Nohara, Sci. Rep. \textbf{3}, 1478 (2013).
\bibitem{LandC_2010} M. D. Lumsden and A. D. Christianson, J. Phys.: Condens. Matter \textbf{22}, 203203 (2010).
\bibitem{Dai_2012} P. Dai, J. Hu and E. Dagotto, Nat. Phys. \textbf{8}, 710 (2012).
\bibitem{Yildirim_2009} T. Yildirim, Phys. Rev. Lett. \textbf{102}, 037003 (2009).
\bibitem{Colonna_2011} N. Colonna, G. Profeta, A. Continenza and S. Massidda, Phys. Rev. B \textbf{83}, 094529 (2011).
\bibitem{Tomic_2012} M. Tomi\'{c}, R. Valent\'{\i} and H. O. Jeschke, Phys. Rev. B \textbf{85}, 094105 (2012).
\bibitem{Widom_2013} M. Widom and K. Quader, arXiv:1207.4550v1, (2012).
\bibitem{Diallo_2010} S. O. Diallo, D. K. Pratt, R. M. Fernandes, W. Tian, J. L. Zarestky, M. Lumsden, T. G. Perring, C. L. Broholm, N. Ni, S. L. Bud'ko, P. C. Canfield, H.-F. Li, D. Vaknin, A. Kreyssig, A. I. Goldman, and R. J. McQueeney, Phys. Rev. B \textbf{81}, 214407 (2010).
\bibitem{Pratt_2009} D. K. Pratt, Y. Zhao, S. A. J. Kimber, A. Hiess, D. N. Argyriou, C. Broholm, A. Kreyssig, S. Nandi, S. L. Bud'ko, N. Ni, P. C. Canfield, R. J. McQueeney, and A. I. Goldman, Phys. Rev. B
\textbf{79}, 060510(R) (2009).
\bibitem{Loong_1987} C. -K. Loong, B. H. Grier, S. M. Shapiro, J. M. Lawrence, R. D. Parks and S. K. Sinha, Phys. Rev. B \textbf{35}, 3092 (1987).
\bibitem{Wang_2013} Meng Wang, Chenglin Zhang, Xingye Lu,  Guotai Tan, Huiqian Luo, Yu Song, Miaoyin Wang, Xiaotian Zhang, E. A. Goremychkin, T. G. Perring, T. A. Maier, Zhiping Yin, Kristjan Haule, Gabriel Kotliar, and Pengcheng Dai, arXiv:1303.7339v1, (2013).
\bibitem{Abernathy_2012} D. L. Abernathy, M. B. Stone, M. J. Loguillo, M. S. Lucas, O. Delaire O, X. Tang X, J. Y. Y. Lin and B. Fultz, Rev. Sci. Instrumen. \textbf{83}, 15114 (2012).
\bibitem{Ran_2011} S. Ran, S. L. Bud'ko, D. K. Pratt, A. Kreyssig, M. G. Kim, M. J. Kramer, D. H. Ryan, W. N. Rowan-Weetaluktuk, Y. Furukawa, B. Roy, A. I. Goldman, and P. C. Canfield, Phys. Rev. B \textbf{83}, 144517 (2011).
\bibitem{Gati_2012} E. Gati, S. K{\"o}hler, D. Guterding, B. Wolf, S. Knöner, S. Ran, S. L. Bud'ko, P. C. Canfield, and M. Lang Phys. Rev. B \textbf{86}, 220511 (2012)
\bibitem{Tucker_2012} G. S. Tucker, D. K. Pratt, M. G. Kim, S. Ran, A. Thaler, G. E. Granroth, K. Marty, W. Tian, J. L. Zarestky, M. D. Lumsden, S. L. Bud'ko, P. C. Canfield, A. Kreyssig, A. I. Goldman and R. J. McQueeney, Phys. Rev. B \textbf{86}, 020503(R) (2012).
\bibitem{Coldea_2004} R. Coldea, MSLICE: A Data Analysis Programme for Time-of-Flight Neutron Spectrometers (2004).



\end{thebibliography}
\end{document}